\documentclass[11pt]{article}

\usepackage{fullpage}
\usepackage{url,fancyheadings,amssymb}

\newcommand{\halmos}{\hspace*{\fill}\rule{1ex}{1.4ex}}
\makeatletter
\def\newproof#1{\@nprf{#1}}
\def\@nprf#1#2{\expandafter\@ifdefinable\csname #1\endcsname
\global\@namedef{#1}{\@prf{#1}{#2}}\global\@namedef{end#1}{\@endproof}}
\def\@prf#1#2{\@beginproof{#2}{\csname the#1\endcsname}\ignorespaces}
\def\@beginproof#1{\rm \trivlist \item[\hskip \labelsep{\bf #1: }]}
\def\@endproof{\halmos \endtrivlist}
\makeatother
\newproof{prf}{Proof}
\newproof{proof}{Proof}
\newproof{sketch}{Proof (sketch)}

\newtheorem{theorem}{Theorem}[section]
\newtheorem{lemma}[theorem]{Lemma}

\newtheorem{proposition}[theorem]{Proposition}
\newtheorem{Cor}[theorem]{Corollary}
\newtheorem{Definition}[theorem]{Definition}

\newcommand{\signal}{c}

\newcommand\iw{{\sc norm\_estimation}}
\newcommand\pehh{{\sc private\_Euclidean\_heavy\_hitters}}

\newcommand\pss{{\sc private-sample-sum}}

\newcommand{\text}[1]{\mbox{#1}}

\newcommand\opt{{\rm opt}}

\newcommand{\poly}{\mathop{\rm poly}}

\newcommand{\prot}{\pi}

\newcommand{\cind}{\stackrel{c}{\equiv }}

\newcommand{\view}{{\sf view}}
\newcommand{\out}{{\sf output}}

\newcommand{\xvec}{{\bf x}}

\renewcommand{\paragraph}[1]{\vspace{-.2truecm} \par \bigskip \noindent{\bf
#1. }}

\newlength{\protocolwidth}
\newcommand{\pprotocol}[5]{
{\begin{figure}[#4]
\begin{center}
\fbox{
        \small
        \hbox{\quad
        \begin{minipage}{0.90\textwidth}
    \begin{center}
    {\bf #1}
    \end{center}
        #5
        \caption{\label{#3} #2}
        \end{minipage}
        \quad}
        }
\end{center}
\end{figure}
} }

\newcommand{\protocol}[4]{
\pprotocol{#1}{#2}{#3}{htbp}{#4}
}

\renewcommand{\S}{{\cal S}}

\newcommand{\eqref}[1]{Eq.~(\ref{#1})}

\newcommand\nerr[1]{{\left\|{#1}\right\|}}
\newcommand\serr[1]{{\left\|{#1}\right\|_{{}\sim{}}}}
\newcommand\serrsq[1]{{\left\|{#1}\right\|\rlap{${}^2$}}_{{}\sim{}}}

\newcommand\supp{\mathop{\rm supp}}

\makeatletter
\def\@fnsymbol#1{\ensuremath{\ifcase#1\or *\or \dagger\or \ddagger\or
   \mathchar "278\or \mathchar "27B\or \|\or **\or \dagger\dagger
   \or \ddagger\ddagger \else\@ctrerr\fi\relax}}
\makeatother

\date{}

\title{Private Approximate Heavy Hitters}

\author{Martin J.\ Strauss\thanks{Departments of Math and EECS, University of
Michigan, Ann Arbor, MI 48109 USA {\tt martinjs@umich.edu}.  Supported
in part by NSF grant DMS-0354600.}
\and
Xuan Zheng\thanks{Department of EECS, University of
Michigan, Ann Arbor, MI 48109 USA {\tt xuanzh@eecs.umich.edu}.  Supported
in part by NSF grant DMS-0354600.}
}

\begin{document}

\maketitle

\rhead{\em Draft of \today.  Please do not distribute.} \lhead{}
\setlength{\headrulewidth}{0pt} \rhead{} \lhead{}
\setlength{\headrulewidth}{0pt} \cfoot{\em TCC 2007 submission,
\today.  } \thispagestyle{fancy}

\begin{abstract}
We consider the problem of private computation of approximate Heavy
Hitters.  Alice and Bob each hold a vector and, in the vector sum,
they want to find the $B$ largest values along with their indices.
While the exact problem requires linear communication, protocols in
the literature solve this problem approximately using polynomial
computation time, polylogarithmic communication, and constantly many
rounds.  We show how to solve the problem {\em privately} with
comparable cost, in the sense that nothing is learned by Alice and Bob
beyond what is implied by their input, the ideal top-$B$ output, and
goodness of approximation (equivalently, the Euclidean norm of the
vector sum).  We give lower bounds showing that the Euclidean norm
must leak by any efficient algorithm.
\end{abstract}

\setcounter{page}{0}
\thispagestyle{empty}
\newpage

\section{Introduction}
\label{sec:intro}

Secure and private multiparty computation has been studied for several
decades, starting with~\cite{Yao82,BGW88}.  Any protocol
for computing a function of several inputs can be converted,
gate-by-gate, to a {\em private} protocol, in which no party learns
anything from the protocol messages other than what can be deduced
from the function's input/output relation.  The computational overhead
is at most polynomial in the size of the inputs.

In recent years, however, input sizes in many problems have grown to the point where
``polynomial computational overhead'' is too coarse a measure; both
computation and communication should be minimized.  For example,
absent
privacy concerns, applications may require that a protocol uses at
most polylogarithmic communication.  General-purpose secure multiparty
computation may blow up communication exponentially, so additional
techniques are needed.  In one theoretical approach, individual
protocols are designed for functions of interest such as database
lookup (the {\em private information retrieval} problem~\cite{CGKS95,KO97,CMS99}) and
building decision trees~\cite{LP02}.  Another approach, the
breakthrough~\cite{NN01}, converts any protocol into a private one
with little communication blowup, but imposes a computational blowup
that may be exponential.

The approach we follow, which was introduced in~\cite{FIMNSW06}, is to
substitute an
approximate function for the desired function.  Many functions of
interest have good approximations that can be computed efficiently
both in terms of computation and communication.  A caveat is that the
traditional definition of privacy is no longer appropriate.  Instead,
a protocol $\pi$ computing an approximation $\widetilde f$ to a
function $f$ is a
private approximation protocol~\cite{FIMNSW06} for $f$ if
\begin{itemize}
\item $\pi$ is a private protocol
for $\widetilde f$ in the traditional sense that the messages of $\pi$
leak nothing beyond what is implied by inputs and $\widetilde f$, {\em
  and},
\item the output $\widetilde f$ leaks nothing beyond what is implied
  by inputs and $f$.
\end{itemize}
Several examples were given
in~\cite{FIMNSW06}.  Another important example, crucial to this
article and the first non-trivial example to achieve polylogarithmic
communication and polynomial computation, was given in~\cite{IW06}.
There, Alice and Bob have vectors $a$ and $b$ of length $N$, taking
integer values in the range $[-M,M]$.  Their goal is to approximate the
Euclidean norm of the sum, $\nerr{a+b}_2$.  The authors show how to
compute an estimate $\serr{a+b}$ such that, if $k$ is a security
parameter,
\begin{itemize}
\item $\frac1{1+\epsilon}\nerr{a+b}_2\le\serr{a+b}\le \nerr{a+b}_2$.
\item The protocol requires $\poly(k\log(M)N/\epsilon)$ local computation,
  $\poly(k\log(M)\log(N)/\epsilon)$ communication, and $O(1)$ rounds.
\item No party learns more from the protocol messages than can be
  deduced from the approximate output
  $\serr{a+b}$ and the relevant party's input, and no party learns
  more from the output $\serr{a+b}$ than can be deduced from the exact
  output $\nerr{a+b}_2$.
\end{itemize}
We will make use of this result.

\subsection{Our Results}

Each of two parties has a vector, $a$ and $b$, and they want a summary
for the vector sum $c=a+b$.  First, we consider the Euclidean approximate
heavy hitters problem, in which there is a parameter, $B$, and the
players ideally want $c_\opt$, the $B$ largest terms in $c$, {\em
  i.e.}, the $B$ biggest values together with the corresponding
indices.  Unfortunately, finding $c_\opt$ exactly requires linear
communication.
Instead, the players use polylogarithmic communication (and polynomial
work and $O(1)$ rounds) to output a vector $\widetilde c$ with
$\nerr{\widetilde c-c}_2\le(1+\epsilon)\nerr{c_\opt-c}_2$.  In our
protocol, the players learn nothing more than what can be deduced from
$c_\opt$ and $\nerr{c}_2$.  (We discuss below the significance of
leaking $\nerr{c}_2$.)  We can immediately use this result as black
box for approximate sparse representations
over any orthonormal basis such as wavelet or Fourier, with similar
costs.  We can also use the result as a black
box for taxicab approximate heavy hitters, {\em i.e.}, finding
$\widetilde c$ with $\nerr{\widetilde
  c-c}_1\le(1+\epsilon)\nerr{c_\opt-c}_1$, leaking $c_\opt$ and
$\nerr{c}_2$.

In the basic result, we give an at-most-$B$-term representation that
is nearly as good (in the Euclidean sense) as the best $B$-term
representation and leaks no more than the best $B$-term representation
{\em and the Euclidean norm}.  Leaking the Euclidean norm represents a
weaker result than not leaking the Euclidean norm, but (i) leaking
$\nerr{c}_2$ is necessary in some circumstances and (ii) computing or
approximating
$\nerr{c}_2$ is desirable in some circumstances.  First, we give a
straightforward lower bound showing that, for some (reasonable) values
of parameters $M,N,\ldots$, computing $\widetilde c$ leaking only
$c_\opt$ requires $\Omega(N)$ communication.  In fact, for some
(artificial) classes of inputs, $\Omega(N)$ communication is needed
unless $\nerr{c}_2$ itself is not only potentially leaked, but
actually computed exactly.  On the other hand, one
can regard the Euclidean norm as
semantically interesting, so that we can regard the top $B$ terms {\em
  together with the Euclidean norm} as a compound, extended summary.
In particular, since $\widetilde c$ is computed, leaking $\nerr{c}_2$
is equivalent to leaking
$\nerr{c}_2^2-\nerr{\widetilde c}_2^2=\nerr{\widetilde c-c}_2^2$, {\em
  i.e.}, the error in our
representation, which is a useful and common thing to want to
compute.  Our protocol indeed can be modified to output an
approximation $\serr{\widetilde c-c}$ with
$\nerr{\widetilde c-c}_2\le\serr{\widetilde c-c}
\le(1+\epsilon)\nerr{\widetilde c-c}_2$, so we can regard the protocol
as solving two cascaded approximation problems: find a
near-best representation $\widetilde c$, then find an approximation
$\serr{\widetilde c-c}$ to $\nerr{\widetilde c-c}_2$.  It is natural
to expect a protocol for $\widetilde c$ to leak $c_\opt$ and a
protocol for $\serr{\widetilde c-c}$ to leak
$\nerr{\widetilde c-c}_2$; while lower bounds prevent that, we can
compute $\widetilde c$ and $\serr{\widetilde c-c}$ {\em
  simultaneously} and guarantee that, {\em overall}, we leak only
$c_\opt$ and $\nerr{\widetilde c-c}_2$.

We give a result for taxicab heavy hitters that produces an at-most-$B$
term representation that is nearly as good (in the taxicab sense) as
the the best $B$-term representation and leaks no more than the best
$B$-term representation and the {\em Euclidean} norm.  Thus we have
shown that the private Euclidean norm approximation can be used for
non-Euclidean problems.
Finally, we also give a result for other orthonormal bases that
involves little additional
algorithmic or privacy work, but demonstrates that the basic
result can be applied in a variety of interesting applications.
It says that we provide an at-most-$B$ term Fourier representation
that is almost as good (in the Euclidean sense) as the best
$B$-term Fourier representation and leaks no more than the best
$B$-term representation and the Euclidean norm.  The Fourier
basis may be substituted by any orthonormal basis, such as
Hadamard or Wavelet.

\subsection{Related Work}

Other work in private communication-efficient protocols for specific
functions includes the
Private Information Retrieval problem~\cite{CGKS95,KO97,CMS99},
building decision trees~\cite{LP02}, set intersection and
matching~\cite{FNP04}, and $k$'th-ranked element~\cite{AMP04}.

The breakthrough~\cite{NN01} gives a general technique for converting
any protocol into a private protocol with little communication
overhead.  It is not the end of the story, however, because the
computation may increase exponentially.

Work in private approximations include~\cite{FIMNSW06} that introduced
the notion as a conference paper in 2001 and gave several protocols.
Some negative results were given in~\cite{HKKN01} for approximations
to NP-Hard functions; more on NP-hard search problems appears
in~\cite{BCNW06}.
Recently, \cite{IW06} gives a private
approximation to the Euclidean norm that is central to our paper.

Statistical work such as~\cite{CDMSSW05} also addresses approximate
summaries over large databases, but differs from our work in many
parameters, such as the number of players and the allowable
communication.

There are many papers that address the Heavy Hitters problem and
sketching in general, in a
variety of contexts.  Many of the needed ideas can be seen in
\cite{KM91} and other important papers
include~\cite{AMS96,AGMS99,GGIKMS02,CM03:Whats-Hot}.

\subsection{Organization}

This paper is organized as follows.  In Section~\ref{sec:prelim}, we
give preliminaries.  In Section~\ref{sec:algo}, we present our main
result.  In Section~\ref{sec:lb}, we present lower bounds.

\section{Preliminaries}
\label{sec:prelim}

\subsection{Parameters and Notation}

Fix parameters $N,M,B,k,\epsilon$.  We will consider two players,
Alice and Bob, who will have inputs, $a$ and $b$ respectively, that
are vectors of length $N$ taking integer values in the range $-M$ to
$+M$.  Throughout, we will be interested in summaries of size $B$ for
the vector $c=a+b$.  For example, in the main result, we are
interested ideally in the largest $B$ terms of $c$.
A vector $c$ is written
$c=(c_0,c_1,c_2,\ldots,c_{N-1})=\sum c_j\delta_j$, where $j$ is an
{\em index}, $c_j$ is a {\em value}, $\delta_j$ is the vector that is
1 at index $j$ and 0 elsewhere, and $c_j\delta_j$, which can be
implemented compactly and equivalently written as the pair $(j,c_j)$,
is a {\em term}, in which $c_j$ is the {\em coefficient}.

We compare terms by the {\em magnitudes} of their coefficients,
braking ties by the indices.  That is, we will say that
$(j,c_j)<(k,c_k)$ if $|c_j|<|c_k|$ or both $|c_j|=|c_k|$ and $j<k$.
Thus all terms are strictly comparable.  A heavy hitter summary is an
expression of the form $\sum_{i\in\Lambda} \eta_i\delta_i$.  If
$|\Lambda|$ must be at most $B$, then the best heavy hitter summary
$c_\opt$ for a vector $c$ occurs where $\{(i,\eta_i):i\in\Lambda\}$
consists of the $B$ largest terms.

The Euclidean norm of $a$ is $\nerr{a}_2=\sqrt{\sum_i
  a_i^2}$ and the taxicab norm is $\nerr{a}_1=\sum_i
|a_i|$.  The {\em support} $\supp(a)$ of a vector $a$ is the set of
indices where $a$ is non-zero, $\{i:a_i\ne 0\}$.

The parameter $\epsilon$ is a distortion parameter.  We will guarantee
summaries whose error is at most the factor $(1+\epsilon)$ times the
error of the best possible summary.

The parameter $k$ is a security and failure probability parameter.
Algorithms will be expected to succeed except with probability
$2^{-k}$ and $2^{-k}$ will serve as an upper bound for the allowable
statistical distance between indistinguishable distributions.

We will be interested in protocols that use
communication $\poly( B, \log(N), k, \log(M), 1/\epsilon)$,
local computation $\poly( B, N, k, \log(M), 1/\epsilon)$, and
number of rounds that is constant.

\subsection{Approximate Data Summaries}

In the heavy hitters problem, we are given parameters $B$ and $N$ and
the goal is to find the $B$ largest terms in a vector $c$ of length
$N$. We will be interested in two approximate versions, parametrized
also by $\epsilon$.  In the approximate heavy hitters
problem, we want a summary $\widetilde c=\sum_{i\in\Lambda}
\eta_i\delta_i$ such that
$\nerr{\widetilde c-c} \le (1+\epsilon)\nerr{c_\opt-c}$,
where the norms are, respectively, 2-norms (in the Euclidean
approximate heavy hitters problem) and 1-norms  (in the taxicab
approximate heavy hitters problem).

In order to describe previous algorithms that are relevant to us,
we first need some definitions.  Fix a vector $c =
(c_0,c_1,c_2,\ldots,c_{N-1}) = \sum_{0\le i<N} c_i\delta_i$, whose
terms are
$t_0=(0,c_0),t_1=(1,c_1),\ldots,t_{N-1}=(N-1,c_{N-1})$.  Suppose the
sequence $i'_0,i'_1,\ldots$ is a decreasing rearrangement of $c$,
{\em i.e.}, $t_{i'_0}>t_{i'_1}>\cdots>t_{i'_{N-1}}$.

\begin{Definition}\label{def:sig}
(Significant index.)
Let $I\subseteq[0,N)$ be a set of indices containing $i$.  Then $i$ is
a {\em ($I, \theta$)-significant index} for $c$ if and only if
$c_i^2\geq \theta\sum_{j\in I}|c_j|^2$.
\end{Definition}
That is, an index is signficant if the corresponding value is large
compared with all the values.  In some of the algorithms below, we
will find the largest term (if it is sufficiently large), subtract it
off, then recurse on the residual signal.  This motivates the
following definitions.

\begin{Definition}\label{def:sigset}
(Significant index set.)
Let $I\subseteq[0,N)$ be a set of $m$ indices containing $i$.  Then
  $I$ is a {\em
    $\theta$-significant index set} for $c$ if and only if
 $\forall
j=1\cdots m$, $t_{i'_j}$ is a
($[0,N)\backslash\{i'_1,\cdots,i'_{j-1}\},\theta$)-significant
index.
\end{Definition}
That is, in a significant index set for $c$, the largest
term has a significant index; after removing the largest term, the new
largest term has a significant index, etc.
Note that there can be more than one $\theta$-significant
index set for a given vector.

\begin{Definition}\label{def:qualify}
(Qualified index set.)  Fix parameters $\ell$ and $\theta$.
The set $Q=\{i'_0,i'_1,\ldots,i'_{m-1}\}$ is a {\em ($\ell,
\theta$)-qualified index set} for $c$ if and only if
\begin{itemize}
\item $m\le\ell$,
\item $\{i'_0,i'_1,\ldots,i'_{m-1}\}$ is a $\theta$-significant index set, and
\item $\{i'_0,i'_1,\ldots,i'_{m-1},i'_{m}\}$ is NOT a $\theta$-significant index set.
\end{itemize}
\end{Definition}
That is, a qualified index set consists of the largest possible
length $m$ for a prefix of $i'_0,i'_1,\ldots,i'_{m-1}$ such that, for each
$j<m$, we have $c_{i'_j}^2\ge\theta(c_{i'_j}^2 + c_{i'_{j+1}}^2 +
c_{i'_{j+2}}^2 + \cdots+c_{N-1}^2)$.  In particular, if the terms
happen to be in decreasing order to begin with, {\em i.e.}, if
$|c_0|>|c_1|>\cdots$, then a qualified index set is
$\{0,1,2,\ldots,m-1\}$ for the largest $m$ such that, for each
$j<m$, we have $c_j^2\ge \theta(c_j^2+c_{j+1}^2+c_{j+2}^2+\cdots
c_{N-1}^2)$.

Note that for each $\ell,\theta$, and vector $c$, there is only one
($\ell, \theta$)-qualified index set for $c$.  We use
$Q_{c,\ell, \theta}$ to denote it.  We sometimes write
$Q_{\ell,\theta}$ when $c$ is understood.

The following are straightforward.
\begin{proposition}\label{prop:qualify}
For any $\theta_1<\theta_2$, $Q_{\ell, \theta_2}$ set is a subset of
$Q_{\ell, \theta_1}$.
\end{proposition}

\begin{proposition}\label{prop:qualisgood}
Fix parameters $N,M,B,k,\epsilon$ and vector $c$ as above.  If
$\widetilde c = \sum_{i\in Q_{c,B,\frac{\epsilon}{B(1+\epsilon)}}}
c_i\delta_i$, then $\nerr{\widetilde
c-c}_2^2\le(1+\epsilon)\nerr{c_\opt-c}_2^2$.
\end{proposition}
\begin{proof}
Assume without loss of generality that $|c_0|>|c_1|>\cdots$ and let
$q=|Q_{c,B,\frac{\epsilon}{B(1+\epsilon)}}|$.  If $q=B$, then
$\widetilde c=c_\opt$ and we are done.  Otherwise we have
\begin{eqnarray*}
\nerr{\widetilde c-c}_2^2
&  =  & \sum_{q\le i<B} |c_i|^2  + \nerr{c_\opt-c}_2^2\\
& \le & B|c_q|^2  + \nerr{c_\opt-c}_2^2\\
& \le & \frac{\epsilon}{1+\epsilon}\nerr{\widetilde c-c}_2^2 +
        \nerr{c_\opt-c}_2^2,
\end{eqnarray*}
whence
\[\left(1-\frac{\epsilon}{1+\epsilon}\right)\nerr{\widetilde c-c}_2^2
\le\nerr{c_\opt-c}_2^2.\]
The result follows.
\end{proof}

The algorithms below will work from a linear {\em sketch} of a vector.
\begin{Definition}\label{def:sketch}
(Sketch of a vector.)  Given a vector $c$, a {\em linear sketch} of
  $c$ is $Rc$, where $R$ is a random matrix generated from a
  prescribed distribution, called the {\em measurement matrix}.
\end{Definition}

In our case, as is typical, the matrix $R$ will be a pseudorandom
matrix, that can be generated from a short pseudorandom seed.  We will
use sketching for the \iw\ protocol, in which the generator needs to be
secure against small space, and a different measurement matrix in the
the non-private Euclidean Heavy Hitters protocol, where, {\em
e.g.}, pairwise independence suffices for the pseudorandom number
generator.

An algorithm in connection with the Euclidean approximate heavy hitter
problem satisfying the following is known:
\begin{theorem}
Fix parameters $N,M,B,k,\epsilon$ as above.  Fix
$\theta\ge\poly(\log(N),\log(M),B,k,1/\epsilon)^{-1}$.  There is a distribution
on sketch matrices $R$ and a corresponding algorithm that, from $R$
and sketch $Rc$ of a vector $c$, outputs a superset of
$Q_{c,B,\theta}$, in time $\poly(\log(N),\log(M),B,k,1/\epsilon)$.
\end{theorem}
In particular, the number or rows in $R$ and the size of the output
is bounded by the expression
$\poly(\log(N),\log(M),B,k,1/\epsilon)$ in accordance with the time bound on the
algorithm.

Note that the algorithm returns a superset of $Q_{c,B,\theta}$ but
that even $Q_{c,B,\theta}$ itself suffices for a good
approximation.

\begin{proof}[sketch]
One such algorithm is as follows.  As in~\cite{GGIKMS02}, one can
estimate $c_i$ by $\widetilde c_i=\delta_i^T R^T Rc\pm
(\epsilon/B)\nerr{c}_2$ except with small probability, where $R$ is
a $\pm1$-valued matrix with $\poly(\log(N),B,1/\epsilon)$
independent rows, each of which is a pairwise independent family. By
repeating $O(k)$ times and taking a median, one can drive down the
failure probability to $2^{-k}$.  As in~\cite{GGIKMS02}, one need not
estimate all the terms; rather, in time
$\poly(\log(N),\log(M),B,k,1/\epsilon)$, one
can find a set $I$ of indices that includes all terms with magnitude
at least $\theta\nerr{c}_2$ (and possibly other terms).  By adjusting
parameters, one can estimate such $c_i$ well enough as $\widetilde
c_i$ so that $|\widetilde c_i-c_i|^2\le (\epsilon/B)\nerr{c}_2^2$.
To get a superset of a qualified set, subtract off the approximation
to $c_i\delta_i$ and repeat as long as new $c_i$ (or better
approximations to old $c_i$) are found that are large compared with
the residual vector.  At most $O(\log(MN))$ repetitions are needed
since, after $O(\log(MN))$ repetitions, we have reduced
$\nerr{c}_2^2$ from its initial value of at most $M^2N$ to its least
possible positive value of 1.
\end{proof}

\subsection{Privacy}\label{subsec:private}

Secure multiparty computation allows two or more parties to
evaluate a specified function of their inputs while hiding their
inputs from each other. We work in the semi-honest model, which
assumes that the adversary is passive and can't modify the
behavior of corrupted parties. In particular, the computation is
only concerned with the information learned by the adversary, and
not with the effect misbehavior may have on the protocol's
correctness.

We briefly review private two-player protocols in the
semi-honest model.  A
two-party computation task is specified by a (possibly randomized)
mapping $g$ from a pair of inputs $(a,b)\in\{0,1\}^*\times
\{0,1\}^*$ to a pair of outputs $(c,d)\in\{0,1\}^*\times
\{0,1\}^*$. Let $\prot=(\prot_A,\prot_B)$ be a two-party protocol
computing $g$. Consider the probability space induced by the
execution of $\prot$ on input $\xvec=(a,b)$ (induced by the
independent choices of random inputs $r_A,r_B$). Let
$\view^\prot_A(\xvec)$ (resp., $\view^\prot_B(\xvec)$) denote the
entire view of Alice (resp., Bob) in this execution, including her
input, random input, and all messages she has received.  Let
$\out^\prot_A(\xvec)$ (resp., $\out^\prot_B(\xvec)$) denote
Alice's (resp., Bob's) output. Note that the above four random
variables are defined over the same probability space.  Two
distributions (or ensembles) $\mathcal{D}_1$ and $\mathcal{D}_2$ are
said to be {\em computationally indistinguishable} with security
parameter $k$, $\mathcal{D}_1\cind\mathcal{D}_2$, if, whenever
$X_1\sim\mathcal{D}_1$ and $X_2\sim\mathcal{D}_2$ and for any function
$C$ having a circuit of size at most $2^k$, we have
then $|\Pr(C(X_1)=1) - \Pr(C(X_2)=1)|\le 2^{-k}$.

\begin{Definition}
\label{def-priv} Let $X$ be the set of all valid inputs
$\xvec=(a,b)$. A protocol $\prot$ is a {\em private protocol
computing $g$} if the following properties hold:
\begin{description}
\item[Correctness.] The joint outputs of the protocol are
distributed
   according to $g(a,b)$. Formally,
\[ \{(\out^\prot_A(\xvec),\out^\prot_B(\xvec))\}_{\xvec\in X} \equiv
  \{(g_A(\xvec),g_B(\xvec))\}_{\xvec\in X},\]
where $(g_A(\xvec),g_B(\xvec))$ is the joint distribution of the
outputs of $g(\xvec)$.
\item[Privacy.] There exist  probabilistic
polynomial-time algorithms $\S_A,\S_B$, called {\em simulators},
such that:
\begin{eqnarray*}
\{(\S_A(a,g_A(\xvec)),g_B(\xvec)) \}_{\xvec=(a,b)\in X}
\cind  \{(\view^\prot_A(\xvec),\out^\prot_B(\xvec))\}_{\xvec\in X} \\
\{(g_A(\xvec),\S_B(b,g_B(\xvec))\}_{\xvec=(a,b)\in X} \cind
\{(\out^\prot_A(\xvec),\view^\prot_B(\xvec))\}_{\xvec\in X}
\end{eqnarray*}
\end{description}
\end{Definition}

There are efficient general techniques:
\begin{proposition}\label{prop:share}(General-Purpose Secure
  Multiparty Computation (SMC)~\cite{Yao82})
Two parties holding inputs $x$ and
$y$ can privately compute any circuit $C$ with communication and
computation
$O(k(|C|+|x|+|y|))$, where $k$ is a security parameter, in $O(1)$ rounds.
\end{proposition}

Private approximation requires further discussion.

\begin{Definition}[Private Approximation Protocol~(\cite{FIMNSW06})]
\label{def:pap}
A two-party protocol $\pi$ is a private approximation protocol for a
deterministic, common-output function $g$ on inputs $a$ and $b$ if
$\pi$ computes a
(possibly randomized) approximation $\widetilde g$ to $g$ such
that
\begin{itemize}
\item $\widetilde g$ is a good approximation to $g$ (in the
  appropriate sense)
\item $\pi$ is a private protocol for $\widetilde g$ in the
  traditional sense.
\item (Functional Privacy.)  There exists a probabilistic
polynomial-time simulator $\S$ such that:
\[\{\S(g(\xvec)) \}_{\xvec=(a,b)\in X} \cind \widetilde g(\xvec).\]
\end{itemize}
\end{Definition}
In our case, $g(a,b)$ will formally be the pair
$(c_\opt,\nerr{c}_2)$ and $\widetilde g(a,b)$ will be $\widetilde c$.
We will informally say that we ``approximate $c_\opt$ leaking only $c_\opt$
and $\nerr{c}_2$,'' since there is a simulator that takes $c_\opt$
and $\nerr{c}_2$ as input and simulates the approximate output
$\widetilde c$ and the protocol messages.  Equivalently, one could
define $g(a,b)$ to be the pair $(c_\opt,\nerr{c_\opt-c})$ and define
$\widetilde g(a,b)$ to be the pair $(\widetilde c,\serr{\widetilde
  c-c})$, where $\serr{\cdot}$ is an approximation to the Euclidean
norm (see below).

In our case of a deterministic function to be output to both Alice and
Bob, a (weakly) equivalent definition is as follows, known as the
``liberal'' definition in~\cite{FIMNSW06}:
\begin{Definition}\label{def:pap-liberal}
A two-party protocol $\pi$ is a private approximation protocol for a
deterministic, common-output function $g$ on inputs $a$ and $b$ in the
{\em liberal sense} if
$\pi$ computes a
(possibly randomized) approximation $\widehat g$ to $g$ such
that
\begin{itemize}
\item $\widehat g$ is a good approximation to $g$ (in the
  appropriate sense)
\item There exists a probabilistic polynomial-time simulators $\S_A$
  and $\S_B$ such that:
\begin{eqnarray*}
\{\S_A(a,g(\xvec))\}_{\xvec=(a,b)\in X} & \cind &
    \{\view^\prot_A(\xvec)\}_{\xvec\in X} \\
\{\S_B(b,g(\xvec))\}_{\xvec=(a,b)\in X} & \cind &
    \{\view^\prot_B(\xvec)\}_{\xvec\in X} \\
\end{eqnarray*}
\end{itemize}
\end{Definition}

Roughly speaking, the equivalence is as follows.  Suppose there
are simulators in the standard definition.  Then, putting
$\widehat g=\widetilde g$, a simulator for the liberal defintion
can be constructed by simulating $\widehat g(a,b)=\widetilde
g(a,b)$ from $g(a,b)$ using the hypothesized simulator for
functional privacy, then simulating Alice's view from $\widehat
g(a,b)$ and $a$ using the hypothesized simulator traditional
simulator for the protocol that computes $\widetilde g$.  In the
other direction, suppose there is a simulator in the liberal
definition.  Let $\tau$ be a transcript of Alice's view except for
input $a$.  (As it turns out, it is not necessary to include $a$
in $\tau$.  If $a$ is much longer than $\tau$---as in our
situation---we want to avoid including $a$ in $\tau$ in order to
keep $\tau$ short.)  Define $\widetilde g=\widehat g.\tau$ to be
$\widehat g$ with $\tau$ encoded into its low-order bits.  We
assume that this kind of encoding into approximations can be
accomplished without significantly affecting the goodness of
approximation; in fact, we will assume that the value represented
does not change at all, even if the ``approximate'' value is
zero---that is, $\tau$ is auxiliary data rather than an actual
part of the value of $\widetilde g$.  Note that a protocol for
$\widehat g$ also serves as a protocol for $\widetilde g$.  It is
trivial to simulate the messages of the protocol given $a$ and
$\widetilde g$.  Use the hypothesized simulator in the liberal
definition to show functional privacy.

We will use the technique of encoding into the low-order bits
in our main result, which, formally, will be proven in the standard
definition.  We remark
that the \iw\ protocol from~\cite{IW06} is presented in the liberal
definition.

We will need the following standard definition.

\begin{Definition}[Additive Secret Sharing]
An intermediate value $x$ of a joint computation is said to be {\em secret
shared} between Alice and Bob if Alice holds $r$ and Bob holds $x-r$,
modulo some large prime, where $r$ is a random number independent of
all inputs and outputs.
\end{Definition}

The Private Sample Sum problem is as follows.
\begin{Definition}[Private Sample Sum]
At the start, Alice holds a vector $a$ of length $N$ and Bob holds a
vector $b$.  Alice and Bob also hold a secret sharing of an index $i$.  At
the end, Alice and Bob hold a secret sharing of $a_i + b_i$.
\end{Definition}
That is, neither the index $i$ nor the value $a_i+b_i$ becomes known
to the parties.  Efficient protocols for this can be found (or can be
constructed immediately from related results) in~\cite{NN01,FIMNSW06},
under various assumptions about the existence of Private Information
Retrieval, such as in~\cite{CMS99}.

\begin{proposition}
There is a protocol \pss\ for the Private Sample Sum problem that requires $\poly(N,k)$
computation, $\poly(\log(N),k)$ communication, and $O(1)$ rounds.
\end{proposition}

Our results also rely on the following protocol from~\cite{IW06}, that
privately approximates the Euclidean norm of the vector sum.

\begin{proposition}\label{prop:l-2}(Private $l_2$ approximation)~\cite{IW06}
Suppose Alice and Bob have integer-valued vectors $a$ and $b$ in
$[-M,M]^N$ and let $c=a+b$.  Fix distortion $\epsilon$ and security
parameter $k$.  There is a protocol \iw\ that computes an
approximation $\serr{c}$ to $\nerr{c}_2$ such that
\begin{itemize}
\item $\frac1{1+\epsilon}\nerr{a+b}_2\le\serr{a+b}\le \nerr{a+b}_2$.
\item The protocol requires $\poly(k\log(M)N/\epsilon)$ local computation,
  $\poly(k\log(M)\log(N)/\epsilon)$ communication, and $O(1)$ rounds.
\item The protocol is a private approximation protocol for $\nerr{c}$
  in the sense of Definition~\ref{def:pap-liberal}.
\end{itemize}
Furthermore, the protocol's only access to $a$ and $b$ is through the
matrix-vector products $Ra$ and $Rb$, where $R$ is a pseudorandom
matrix known to both players.
\end{proposition}

\section{Private Euclidean Heavy Hitters}
\label{sec:algo}

We consider the setting in which Alice has signal $a$ of
dimension $N$, and Bob has signal $b$ of the same
dimension.  Let $c=a+b$. Both parties want to learn a representation
$\widetilde c = \sum_{t\in T_{\rm out}}t$ such that
$\nerr{c-\widetilde c}_2^2\le(1+\epsilon) \nerr{c-c_{\opt}}_2^2$ and
such that at most $c_\opt$ and $\nerr{c}_2$ is revealed.  A protocol is
given in Figure~\ref{fig:ehh}.

\protocol{\pehh}{Protocol for the Euclidean Heavy
  Hitters problem}{fig:ehh}{
\begin{itemize}
\item Known structural parameters: $N,M,B,\epsilon,k$, which determine
  $\theta=\frac{\epsilon}{B(1+\epsilon)}$ and $B'$
\item Individual inputs: vectors $a$ and $b$, of length $N$, with
  integer values in the range $[-M,M]$.
\item Output: With probability at least $1-2^{-k}$, a set $T_{\rm
  out}$ of at most $B$ terms, such that
$\nerr{\signal-\sum_{t\in T_{\rm out}}t}_2^2\le(1+\epsilon)
   \nerr{\signal-\sum_{t\in T_\opt}t}_2^2$.
\end{itemize}
\hrulefill
\begin{enumerate}
\item Exchange pseudorandom seeds (in the clear).  Generate
  measurement matrices $R_1$ and $R_2$.  Alice locally constructs
  sketches $R_1a$ and $R_2a=(R_2^0a, R_2^1a,\ldots R_2^{B-1}a)$,
  where the measurement matrix $R_1$ is used for a non-private
  Euclidean Heavy Hitters and the
  measurement matrix
  $R_2=(R_2^0,R_2^1,\ldots,R_2^{B-1})$ is used for $B$ independent
  repetitions of \iw.  Bob similarly constructs $R_1b$ and $R_2b$.
\item
\label{step:ehh}
  Using general-purpose SMC, do
\begin{itemize}
  \item
  Use an existing (non-private) Euclidean Heavy Hitters protocol to get,
  from $R_1a$ and $R_1b$, a secret-sharing of a
  superset $I$ of $Q_{c,B,\frac{\theta}{1+\epsilon}}$, in which $I$ has exactly
  $B'\le \poly(\log(N),\log(M),B,k,1/\epsilon)$ indices.    (Pad with
  arbitrary indices, if necessary.)
\end{itemize}
\item
\label{step:haveterms}
  Use \pss\ to compute, from $I,a$, and
  $b$, secret-shared values for each index in $I$.  Let $T$
  denote the corresponding set of secret-shared
  terms.  (Both the index and value of each term in $T$ is secret
  shared.)  Enumerate $I$ as $I=\{i_0,i_1,\ldots\}$ with
  $t_{i_0}>t_{i_1}>\cdots$.
\item
\label{step:pursuit}
Using SMC, do
\begin{itemize}
\item for $j=0$ to $B-1$
\begin{enumerate}
  \item From $R_2^j, R_2^ja, R_2^jb, t_0, t_1, \ldots, t_{i_{j-1}}$,
    sketch $r_j=c-(t_{i_0}+t_{i_1}+\cdots+t_{i_{j-1}})$ as
    $R_2^jr_j = (R_2^ja + R_2^jb - R_2^j(t_{i_0}+t_{i_1}+\cdots+t_{i_{j-1}}))$.
  \item use \iw\ to estimate $\nerr{r_j}_2^2$ as
  $\serrsq{r_j}$, satisfying
  $\frac{1}{1+\epsilon}\nerr{r_j}_2^2
   \le \serrsq{r_j}\le \nerr{r_j}_2^2$.
  \item If $|c_{i_j}|^2<\theta\serrsq{r_j}$, break (out of for-loop)
  \item Output $t_j$
\end{enumerate}
\end{itemize}
\item For technical reasons, encode the pseudorandom seeds for $R_1$ and
$R_2$ into the low-order bits of the output or (as we assume here)
  provide $R_1$ and $R_2$ as auxiliary output.
\end{enumerate}
}

\subsection{Analysis}

First, to gain intuition, we consider some easy special cases of the
protocol's operation.  For our analysis, assume that the terms in $c$ are
already positive and in decreasing order, $c_0>c_1>\cdots>c_{N-1}>0$.
We will be able
to find the coefficient value of any desired term, so we focus on the set of
indices.  Let $I_\opt=\{0,1,2,\ldots,B-1\}$ denote the set of indices
for the optimal $B$ terms.  Thus $Q_{c,B,\theta}\subseteq
Q_{c,B,\frac{\theta}{1+\epsilon}}\subseteq I_\opt$ and
$Q_{c,B,\frac{\theta}{1+\epsilon}}\subseteq I$.

The ideal output is $I_\opt$, though any superset of $Q_{c,B,\theta}$
suffices to get an approximation with error at most $(1+\epsilon)$
times optimal.  This includes the set $I\supseteq Q_{c,B,\theta}$
which the algorithm has recovered.  The set $I_B$ of the largest $B$
terms indexed by $I$ contains $Q_{c,B,\theta}$, so $I_B$ is a set of
at most $B$ terms with error at most $(1+\epsilon)$ times optimal.  If
$|Q_{c,B,\theta}|=B$, then $I_B=Q_{c,B,\theta}=I_\opt$, and $I_B$ is a
private and correct output.

The difficulty arises when $|Q_{c,B,\theta}|<B$, in which case some of
$I_B$ may be arbitrary and should not be allowed to leak.  So the
algorithm needs to find a private subset $I_{\rm out}$ with
$Q_{c,B,\theta}\subseteq I_{\rm out}\subseteq I_B$.
The challenge is subtle.
Let $s$ denote $|Q_{c,B,\theta}|$.  If the algorithm knew $s$, the
algorithm could
easily output $Q_{c,B,\theta}$, which is the indices of the top $s$
terms, a correct and private output.
Unfortunately, determining $Q_{c,B,\theta}$ or $s=|Q_{c,B,\theta}|$
requires $\Omega(N)$ communication (see Section~\ref{sec:lb}), so we
cannot hope to find $Q_{c,B,\theta}$ exactly.  Non-private norm
estimation can be
used to find a subset $I_{\rm out}$ with
$Q_{c,B,\theta}\subseteq I_{\rm out}\subseteq
Q_{c,B,\frac{\theta}{1+\epsilon}}\subseteq I_\opt$,
which is correct, but not quite private.  Given $|I_{\rm out}|$, the
contents of $I_{\rm out}\subseteq I_\opt$ are indeed private, but the
{\em size} of $I_{\rm out}$ is, generally, non-private.  Fortunately,
if we use a {\em private} protocol for norm estimation, $|I_{\rm
  out}|$ remains private.  We now proceed to a formal analysis.

\begin{theorem}
Protocol~\pehh\ requires
$\poly(N,\log(M),B,k,1/\epsilon)$ local computation,
$\poly(\log(N),\log(M),B,k,1/\epsilon)$ communication, and
$O(1)$ rounds.
\end{theorem}
\begin{proof}
By existing work, all costs of Steps~1 to~\ref{step:haveterms} are as
claimed.  Now consider Step~\ref{step:pursuit}.  Observe that the
function being computed in Step~\ref{step:pursuit} has inputs and
outputs of size bounded by
$\poly(\log(N),\log(M),B,k,1/\epsilon)$ and takes time polynomial in
the size of its inputs.  In particular, the instances of \iw\ do {\em
  not} start from scratch with a reference to $a$ or $b$; rather, they
pick up from the precomputed short sketches $R_2a$ and $R_2b$.
It follows that this function can be wrapped with SMC,
preserving the computation and communication up to polynomial blowup
in the size of the input and keeping the round complexity to $O(1)$.
\end{proof}

We now turn to correctness and privacy.  Let $I_{\rm out}$ denote the
set of indices corresponding to the
set $T_{\rm out}$ of output terms.

\begin{theorem}\label{thm:correct}
Protocol~\pehh\ is correct.
\end{theorem}

\begin{proof}
The correctness of Steps~\ref{step:ehh} and~\ref{step:haveterms}
follows from previous work.  In Step~\ref{step:pursuit}, we first show
that
$Q_{B,\frac{\epsilon}{B(1+\epsilon)}}\subseteq I_{\rm out}$.

We assume that
$\frac{1}{1+\epsilon}\nerr{r_j}_2^2\leq \serrsq{r_j} \leq
\nerr{r_j}_2^2$ always holds; by Proposition~\ref{prop:l-2}, this
happens with high probability.  Thus,
if $|c_{i_j}|^2\geq
\frac{\epsilon}{B(1+\epsilon)}\nerr{r_j}_2^2$, then
$|c_{i_j}|^2\geq \frac{\epsilon}{B(1+\epsilon)}\nerr{r_j}_2^2\geq
\frac{\epsilon}{B(1+\epsilon)} \serrsq{r_i}$.

By construction,
$Q_{B,\frac{\epsilon}{B(1+\epsilon)}}\subseteq I$.
A straightforward induction shows that, if $j\in
Q_{B,\frac{\epsilon}{B(1+\epsilon)}}$, then iteration $j$ outputs
$t_{i_j}$ and the previous iterations output exactly the set of the $j$
larger terms in $I$.

By Proposition~\ref{prop:qualisgood}, since $I_{\rm out}$ is a
superset of $Q_{B,\frac{\epsilon}{B(1+\epsilon)}}$, if $\widetilde
c=\sum_{j\in
  I_{\rm out}} c_{i_j}\delta_{i_j}$, then $\nerr{\widetilde
  c-c}_2^2\le(1+\epsilon)\nerr{c_\opt-c}_2^2$, as desired.
\end{proof}

Before giving the complete privacy argument, we give a lemma, similar
to the above.  Suppose a set $P$ of indices
is a subset of another set $Q$ of indices.  We will say that $P$
is a {\em prefix} of $Q$ if $i\in P, t_j > t_i$, and $j\in Q$ imply
$j\in P$.
\begin{lemma}
The output set $I_{\rm out}$ is a prefix of
$Q_{B,\frac{\epsilon}{B(1+\epsilon)^2}}$ except with probability $2^{-k}$.
\end{lemma}
\begin{proof}
Note that $Q_{B,\frac{\epsilon}{B(1+\epsilon)^2}}$ is a subset of
$I$ and $Q_{B,\frac{\epsilon}{B(1+\epsilon)^2}}$ is a prefix of
the universe, so $Q_{B,\frac{\epsilon}{B(1+\epsilon)^2}}$ is a
prefix of $I$.  The set $I_{\rm out}$ is also a prefix of $I$.  It
follows that, of the sets $I_{\rm out}$ and
$Q_{B,\frac{\epsilon}{B(1+\epsilon)^2}}$, one is a prefix of the
other (or they are equal).

So suppose, toward a contradiction, that
$Q_{B,\frac{\epsilon}{B(1+\epsilon)^2}}$ is a proper prefix of
$I_{\rm out}$.  Let
$q=\left|Q_{B,\frac{\epsilon}{B(1+\epsilon)^2}}\right|$, so $q$ is the
least number such that $i_q$ is {\em not} in
$Q_{B,\frac{\epsilon}{B(1+\epsilon)^2}}$.  If
the protocol halts before considering $q$, then $I_{\rm out}\subseteq
Q_{B,\frac{\epsilon}{B(1+\epsilon)^2}}$, a contradiction.  So, in
particular, we may assume that $q<B$ (so the for-loop doesn't
terminate).  Then, by definition of
$Q_{B,\frac{\epsilon}{B(1+\epsilon)^2}}$, we
have $|c_{i_q}|^2 <\frac{\epsilon}{B(1+\epsilon)^2}\sum_{j\ge q} |c_{i_j}|^2$.
It follows that
\begin{eqnarray*}
|c_{i_q}|^2
&  <  & \frac{\epsilon}{B(1+\epsilon)^2}\sum_{i\ge q} |c_i|^2\\
&  =  & \frac{\epsilon}{B(1+\epsilon)^2}\nerr{r_q}_2^2\\
& \le & \frac{\epsilon}{B(1+\epsilon)}\serrsq{r_q}.
\end{eqnarray*}
Thus the protocol halts without outputting $t_q$, after outputting
exactly the elements in $Q_{B,\frac{\epsilon}{B(1+\epsilon)^2}}$.
\end{proof}

Finally, we turn to privacy.

\begin{theorem}
Protocol~\pehh\ leaks no more than
  $\nerr{c}_2^2$ and $c_\opt$.
\end{theorem}
\begin{proof}
With the random inputs $R_1$ and $R_2$ encoded into the output, it is
straightforward to show
that Protocol~\pehh\ is a private protocol in the traditional sense
that the protocol messages leak no more than the inputs and outputs.
This is done by composing simulators for \pss\ and SMC.
It remains only to show only that we can simulate the joint
distribution on $(\widetilde c, R_1,R_2)$ given as simulator-input
$c_\opt$ and $\nerr{c}$.  We will show that $R_1$ is indistinguishable
from independent of the joint distribution of $(\widetilde c, R_2)$,
which we will simulate directly.

First, we show that $R_1$ is independent.  Except with probability
$2^{-\Omega(k)}$, the intermediate set
$I$ is a superset of $Q_{B,\frac{\epsilon}{B(1+\epsilon)^2}}$ and the
norm estimation is correct.  In that case, the protocol outputs a
prefix of $Q_{B,\frac{\epsilon}{B(1+\epsilon)^2}}$ and we get
identical output if $I$
is replaced by $Q_{B,\frac{\epsilon}{B(1+\epsilon)^2}}$.
Also, $Q_{B,\frac{\epsilon}{B(1+\epsilon)^2}}$ can be constructed from
$c_\opt$ and $\nerr{c}_2$.  Since the protocol proceeds without
further reference to $R_1$, we have shown that the pair
$(\widetilde c,R_2)$ is indistinguishable from being independent of
$R_1$.  It remains only to simulate $(\widetilde c,R_2)$.

Note that the output $\widetilde c$ does depend non-negligibly on
$R_2$.  If $|c_{i_j}|^2$ is very close to $\theta\nerr{r_j}_2^2$, then
the test $|c_{i_j}|^2<\theta\serrsq{r_j}$ in the protocol may succeed
with probability non-negligibly far from 0 and from 1, depending on
$R_2$, since
the distortion guarantee on $\serrsq{r_j}$ is only the factor
$(1\pm\epsilon)$.

The simulator is as follows.
Assume that the terms in $c_\opt$ are $t_0,t_1,\ldots,t_{B-1}$ with
decreasing order, $t_0>t_1>\cdots>t_{B-1}$.
For each $j\le B$, compute
$E_j=\nerr{c-(t_0+t_1+\cdots+t_{j-1})}_2^2
=\nerr{c}_2^2-\nerr{t_0+t_1+\cdots+t_{j-1}}_2^2$ and then run the~\iw\
simulator on input $E_j$ and $\epsilon$ to get a
sample from the joint distribution $(\widetilde E_j,\overline R_2)$, where
$\widetilde E_j$ is a good estimate to $E_j$.
Our simulator then outputs $t_{i_j}$ if
$|c_{i_j}|^2\ge\frac{\epsilon}{B(1+\epsilon)}\widetilde E_j$, and
halts, otherwise,
following the final for-loop of the protocol.  Call the output of the
simulator $\widetilde s=\sum_j t_{i_j}\delta_{i_j}$.

Again using the fact that a prefix of
$Q_{B,\frac{\epsilon}{B(1+\epsilon)^2}}$ is output, if $j\in
Q_{B,\frac{\epsilon}{B(1+\epsilon)^2}}$, then $i_j=j$; {\em i.e.}, the
$j$'th largest output term is the $j$'th largest overall, so that, if
$j$ is output, we have
$E_j=\nerr{r_j}_2^2$.  Thus $(\widetilde E_j,\overline R_2)$ is distributed
indistinguishably from $(\serrsq{r_j},R_2)$.  The protocol finishes
deterministically using $I$ and $\serrsq{r_j}$ and the simulator
finishes deterministically using
$Q_{B,\frac{\epsilon}{B(1+\epsilon)^2}}$ and $\widetilde E_j$, but,
since the protocol output is identical if $I$ is replaced by
$Q_{B,\frac{\epsilon}{B(1+\epsilon)^2}}$, the
distributions on output $(\widetilde c,R_2)$ of the protocol and
$(\widetilde s,\overline R_2)$ of the simulator are
indistinguishable.
\end{proof}

In summary,
\begin{theorem}\label{thm:main}
Suppose Alice and Bob hold integer-valued vectors $a$ and $b$ in
$[-M,M]^N$, respectively.
Let $B$, $k$ and $\epsilon$ be user-defined parameters.
Let $c=a+b$. Let $T_\opt$ be the set of the largest $B$
terms in $c$.  There is an protocol, taking $a$, $b$, $B$ $k$ and
$\epsilon$ as input, given $T_\opt$ and $\|\signal\|_2^2$, computes a
representation $\widetilde c$ of at most $B$ terms such that:
\begin{itemize}
\item $\nerr{\widetilde c-c}_2\le(1+\epsilon)\nerr{c_\opt -c}_2$.
\item The algorithm uses $\poly(N,\log(M),B,k,1/\epsilon)$ time,
  $\poly(\log(N),\log(M),B,k,1/\epsilon)$ communication, and $O(1)$
   rounds.
\item The protocol succeeds with probability $1-2^{-k}$ and leaks
  only $c_\opt$ and $\nerr{c}_2$ with security parameter $k$.
\end{itemize}
\end{theorem}

\begin{Cor}
With the same hyptotheses and resource bounds, there is a protocol
that computes $\widetilde c$ and an
approximation $\serr{\widetilde c - c}$ to $\nerr{\widetilde c - c}_2$ such
that $\frac1{1+\epsilon}\nerr{\widetilde c - c}_2 \le
\serr{\widetilde c - c} \le \nerr{\widetilde c - c}_2$ and the
protocol leaks only $c_\opt$ and $\nerr{\widetilde c - c}_2$.
\end{Cor}
\begin{proof}
Run the main protocol and output also $\serr{\widetilde c - c}$, which
is computed in the course of the main protocol.  Note that
$\nerr{\widetilde c - c}_2^2=\nerr{c}_2^2 - \nerr{\widetilde c}_2^2$ and
both $\nerr{c}_2$ and $\widetilde c$ are available to the main
simulator (as input and output, respectively), so we can modify the
main simulator to compute $\nerr{\widetilde c - c}_2^2$ as well.
\end{proof}

\subsection{Extension to Taxicab Heavy Hitters}

In this section, we show that our result of Euclidean approximation can be
extended to approximate taxicab heavy hitters.

\begin{lemma}
Let $\widetilde c$ be the output of~\pehh.  If
$\|\signal-\widetilde c\|_2\leq(1+\epsilon)\|\signal-\signal_\opt\|_2$,
then
$\|\signal-\widetilde c\|_1\leq(1+\sqrt{B\epsilon})\|\signal-\signal_\opt\|_1$.
\end{lemma}
\begin{proof}
Let $(i,c_i)$ be the largest term which is not in $Q_{B,
\frac{\epsilon}{B(1+\epsilon)}}$.  {From} Theorem~\ref{thm:main} we
know
$(\sum_{i\le j<B} c_j^2)^{\frac{1}{2}}
<\sqrt{\epsilon}(\sum_{B\le j<N} c_j^2)^{\frac{1}{2}}$.  Using the
fact that
$\frac{1}{\sqrt{|\supp(x)}|}\|x\|_1\leq \|x\|_2\leq\|x\|_1$ for any
signal $x$, we get
\[
\frac{1}{\sqrt{B}}\sum_{i\le j<B} |c_j|
\leq \left(\sum_{i\le j<B} c_j^2\right)^{\frac{1}{2}}
\leq \sqrt{\epsilon}\left(\sum_{B\le j< N} c_j^2\right)^{\frac{1}{2}}
\leq \sqrt{\epsilon}\sum_{B\le j< N} |c_j|.
\]
Thus we have
$\|\signal-\widetilde c\|_1\leq\sum_{i\le j<N} |c_j|
=\sum_{i\le j<B} c_j +\sum_{B\le j<N} |c_j|
=(\sqrt{\epsilon B}+1)\sum_{B\le j<N} |c_j|
=(\sqrt{\epsilon B}+1)\|\signal-\signal_\opt\|_1$.
\end{proof}

Theorem~\ref{thm:l1} follows directly:

\begin{theorem}\label{thm:l1}
Suppose Alice and Bob hold integer-valued vectors $a$ and $b$ in
$[-M,M]^N$, respectively.
Let $B$, $k$ and $\epsilon$ be user\-defined parameters.
Let $c=a+b$. Let $T_\opt$ be the set of the largest $B$
terms in $c$.  There is an protocol, taking $a$, $b$, $M,N,B,k$ and
$\epsilon$ as input, and computes a
representation $\widetilde c$ of at most $B$ terms such that:
\begin{itemize}
\item $\nerr{\widetilde c-c}_1\le(1+\epsilon)\nerr{c_\opt -c}_1$.
\item The algorithm uses $\poly(N,\log(M),B,k,1/\epsilon)$ time,
  $\poly(\log(N),\log(M),B,k,1/\epsilon)$ communication, and $O(1)$
   rounds.
\item The protocol succeeds with probability $1-2^{-k}$ and leaks
  only $c_\opt$ and $\nerr{c}_2$ with security parameter $k$.
\end{itemize}
\end{theorem}

\subsection{Extension to other Orthonormal Bases}

In this section, we consider other orthonormal bases, such as the
Fourier basis.  Alice and Bob hold vectors $a$ and $b$ as before, and
want the $B$ largest Fourier terms---frequencies and corresponding
coefficient values.  The exact problem requires
$\Omega(N)$ communication, so they settle for an approximation,
namely, they want a $B$-term Fourier representation $\widetilde c$
such that $\nerr{\widetilde c-c}_2\le(1+\epsilon)\nerr{c_\opt-c}_2$,
where $c_\opt$ is the best possible $B$-term Fourier representation.

We note that a straightforward generalization of our main result
solves this problem privately and efficiently.  Alice and Bob locally
compute the inverse Fourier transform $F^{-1}a$ and $F^{-1}b$ of their
vectors $a$ and $b$.  Because the Fourier transform is linear,
$x=F^{-1}c=F^{-1}a+F^{-1}b$.  Alice and Bob now want to compute an
approximation to the ordinary heavy hitters for the vector $x$.
Suppose the result is $\widetilde x$.  Then $\widetilde x$ is the
compact collection of Fourier terms and $\widetilde c=F\widetilde x$
is the corresponding approximate
representation of $c$.  By the
Parseval equality, since the Fourier basis is orthogonal, for any $y$,
we have
$\nerr{y}_2=\nerr{Fy}_2=\nerr{F^{-1}y}_2$.  It follows that
$\nerr{\widetilde c-c}_2\le(1+\epsilon)\nerr{c_\opt-c}_2$ if and only if
$\nerr{\widetilde x-x}_2\le(1+\epsilon)\nerr{x_\opt-x}_2$,
so the algorithm is correct when transformed to the Fourier domain.
It also follows that leaking $\nerr{c}_2$ is equivalent to leaking
$\nerr{Fc}_2$, so the algorithm is private when transformed to the
Fourier domain.  Alice and Bob require the additional overhead of
computing a Fourier transform locally, which fits within the overall
budget.

\section{Lower Bounds}\label{sec:lb}

In this Section, we show some lower bounds for problems related to our
main problem, such as computing an approximation to $c_\opt$ without
leaking $\nerr{c}_2$.  The results are straightforward, but we
include them to motivate the approximation and leakage of the
protocols we present.

\begin{theorem}
There is an infinite family of settings of parameters
$M,N,B,k,\epsilon$ such that any protocol that computes the Euclidean
norm exactly on the sum  $c$ of individually-held inputs $a$ and $b$,
uses communication $\Omega(N)$.  Similarly, any protocol that computes
the exact Heavy Hitters or computes
the qualified set $Q_{c,1,1}$ exactly uses communication $\Omega(N)$.
\end{theorem}
\begin{proof}
Consider the set disjointness problem, which requires $\Omega(N)$
communication~\cite{KN97}.  Alice and Bob hold $\{0,1\}$-valued vectors $a$ and
$b$ of length $N$ such that each of $a$ and $b$ has exactly $(N/4)$ 1's
and the supports are either disjoint or intersect in exactly one
index.  The task is to determine the intersection size.  Then, if
$c=a+b$, we have $\nerr{c}_2^2=N/2$ or $\nerr{c}_2^2=N/2+3$, depending
on the size of the intersection, so a protocol for $\nerr{c}_2$ can be
used to solve the set disjointness problem.  Similarly, finding the
one largest heavy hitter solves the set disjointness problem.

Now consider vectors of length $N+1$ in which indices $0$ to $N-1$
directly code an instance of set disjointness as above and index $N$
has a value that is always $\sqrt{N/2+2}$.  Then $|Q_{c,1,1}|=1$ or
$|Q_{c,1,1}|=0$ depending on the norm of indices $0$ to $N-1$, which
requires communication $\Omega(N)$ to determine.
\end{proof}

The above theorem motivates our study of {\em approximate} heavy
hitters, for which there are protocols with exponentially better
communication cost than the exact heavy hitters problem.  The next
theorem motivates leaking the Euclidean norm, by showing that {\em
  any} efficient protocol for the approximate heavy hitters problem
leaks the Euclidean norm on all instances within a class.

\begin{theorem}\label{thm:lb-main}
There is an infinite family of settings of parameters
$M,N,B,k,\epsilon$ such that any protocol that solves the Euclidean
Heavy Hitters
problem on the sum $c$ of individually-held inputs $a$ and $b$,
leaking only $c_\opt$, uses communication $\Omega(N)$.  Furthermore,
for an infinite
class of inputs in which $\nerr{c}_2$ is not constant, any such
protocol either computes $\nerr{c}_2$ or uses
communication $\Omega(N)$.
\end{theorem}
\begin{proof}
Consider vectors $c$ of one of two cases, given by random permutations
of the following vectors:
\[
\left\{
\begin{array}{ll}
(2N,\overbrace{1,1,\ldots,1}^{N/2-1},0,0,\ldots,0), & \mbox{(case 1)}\\
(2N,\overbrace{N,N,\ldots,N}^{N/2-1},0,0,\ldots,0), & \mbox{(case 2).}
\end{array}
\right.
\]
Fix $B=1$ and $\epsilon\gg1/N$.  A correct
protocol finds the top term in case 1.  In case 2, it turns out that
the correctness requirement is vacuous, but, fortunately, the privacy
requirement is useful.
A protocol leaking only
$c_\opt$ must behave indistinguishably in cases 1 and 2 since
$c_\opt$ is the same, so a private protocol reliably finds the the top
coefficient in case 2.  Since a protocol for case 2 can be used to
solve the set disjointness problem, such a protocol uses $\Omega(N)$
bits of communication.  In particular, any protocol either behaves
differently on the two cases---thereby computing $\nerr{c}_2$
for inputs in the union of the two cases---or uses communication
$\Omega(N)$.
\end{proof}

Note that the above theorem also shows that it is impossible in some
cases to solve the approximate {\em taxicab} heavy hitters problem
efficiently without leaking the {\em Euclidean} norm.

Although the class of inputs above is contrived, the (implied)
parameter settings are natural, {\em i.e.},
$\log(M),\log(N),B,k,1/\epsilon$ can be made to be polynomially
related, etc.

\bibliographystyle{plain}

\end{document}